\documentclass[sn-mathphys,Numbered]{sn-jnl}% Math and Physical Sciences Reference Style
\usepackage{graphicx}%
\usepackage{multirow}%
\usepackage{amsmath,amssymb,amsfonts}%
\usepackage{amsthm}%
\usepackage{mathrsfs}%
\usepackage[title]{appendix}%
\usepackage{xcolor}%
\usepackage{textcomp}%
\usepackage{manyfoot}%
\usepackage{booktabs}%
\usepackage{algorithm}%
\usepackage{algorithmicx}%
\usepackage{algpseudocode}%
\usepackage{listings}%
\usepackage{booktabs}

\usepackage{orcidlink}

\theoremstyle{thmstyleone}%
\theoremstyle{thmstyletwo}%

\theoremstyle{thmstylethree}%

\begin{document}

\title{Artificial Intelligence for Scientific Research:
Authentic Research Education Framework}

\author[1,2]{Sergey V Samsonau\thanks{Corresponding author: ssamsonau@gradcenter.cuny.edu}\orcidlink{0000-0002-0835-2970}}
\author[1,2]{Aziza Kurbonova}
\author[1,2]{Lu Jiang}
\author[1,2]{Hazem Lashen}
\author[1,2]{Jiamu Bai}
\author[1,2]{Theresa Merchant}
\author[1,2]{Ruoxi Wang}
\author[1,2]{Laiba Mehnaz}
\author[1,2]{Zecheng Wang}
\author[1,2]{Ishita Patil}

\affil[1]{Tandon School of Engineering, New York University}

\affil[2]{AI for Scientific Research (\href{https://www.aifsr.com/}{www.aifsr.com})}


\abstract{We report a framework that enables the wide adoption of authentic research educational methodology at various schools by addressing common barriers. The guiding principles we present were applied to implement a program in which teams of students with complementary skills develop useful artificial intelligence (AI) solutions for researchers in natural sciences. To accomplish this, we work with research laboratories that reveal/specify their needs, and then our student teams work on the discovery, design, and development of an AI solution for unique problems using a consulting-like arrangement. To date, our group has been operating at New York University (NYU) for seven consecutive semesters, has engaged more than a hundred students, ranging from first-year college students to master's candidates, and has worked with more than twenty projects and collaborators. While creating education benefits for students, our approach also directly benefits scientists, who get an opportunity to evaluate the usefulness of machine learning for their specific needs.}




\maketitle

\section{Introduction}

There are many examples of AI (artificial intelligence) being used in natural sciences: AI for Large Hadron Collider experiments \cite{Castelvecchi2015ArtificialIC}, AI for gravitational wave detection \cite{Chaturvedi2022InferenceOptimizedAA}, AI for microscopy \cite{Kim2019AIpoweredTL}, AI for spectroscopy \cite{Romeu2021CombinedMO, Ho2022BeyondIE}, and AI for electroencephalography \cite{Asghar2021AIIE,Yang2021ACS} to name a few. Report \cite{stevens2020ai} provides a systematic overview of where AI can help scientific research.

While potential benefits are great, applying AI to scientific tasks may be more challenging than applying AI to business cases. Scientific datasets are often unique in the sense that they are not "similar" to many other datasets and thus require a significant effort to adopt existing "typical approaches" developed for generally available datasets and tasks. In addition, in order to even try AI techniques, labs require specialists with a rare combination of skills: people who understand specific areas of science and have practical expertise in AI.

At the same time, there is a large number of students who study machine learning and science and would love to work on these tasks. Machine learning education in classroom settings often revolves around the use of exercises that have already been explored thoroughly by more experienced practitioners, such as building a classifier for the MNIST dataset or addressing datasets published on platforms such as Kaggle. This provides a controllable environment for educators with which they can administer these exercises on the scale of a classroom or cohort. While such exercises can be a beneficial introductory experience for students, the fact that solutions to these exercises have already been made and are readily available means that they are of limited value in providing students with more in-depth expertise in the field, as they are often only expected to replicate prior work done by others. A similar issue exists in the natural sciences, where the majority of labs are taught using a "cookbook approach", in which an instructor provides all the steps which their students have to follow, thus not developing their own experience of the research process.

To overcome the weaknesses of the "cookbook approach", multiple types of 'active learning' education have been developed and shown to improve scientific literacy \cite{Gormally2009EffectsOI}, graduation rates \cite{Rodenbusch2016EarlyEI}, and, potentially, students' well-being \cite{Walkington2022HowDoesEngaging}. One of the types of active learning education family is 'Authentic research education' (AREd), which we will be discussing here. 

In this manuscript, we outline the principles of a framework that allows the design and implementation of novel AREd-type programs and argue that such a framework overcomes common AREd adoption barriers. We demonstrate the use of the principles through the discussion of the program "AI for Scientific Research" (AIfSR) at New York University.  Our focus here is not on the pedagogical outcomes of this particular program but rather on the question of how to simplify the implementation of such programs. Evaluation of AREd programs can be found elsewhere (for example, in \cite{Gormally2009EffectsOI, Rodenbusch2016EarlyEI}).

\section{AREd: definition and implementation barriers}

There is no single widely accepted definition of what an AREd program is. In \cite{Redefining_authentic_research}, researchers surveyed hundreds of biology faculty members to understand better how they conceptualize AREd. Their results provide a useful framework with two common themes present: "Process of Science" and "Novel questions" (the details are presented in table \ref{tab:Definition}). We will use that framework throughout the manuscript. 

In the same work \cite{Redefining_authentic_research}, researchers asked faculty to list barriers to AREd implementation. We will use their summary but will categorize barriers somewhat differently to improve clarity and also will add an additional factor of "sufficient time for project execution". Table \ref{tab:blockers} presents a list of barriers that we will use.

\begin{table*}
  \centering
  \begin{tabular}{p{0.2\textwidth}p{0.65\textwidth}}
    \toprule
    Process of science & A student is engaged in practices built around steps and methods in traditional research (student-generated questions, hypothesis formulation, experimental design, data collection, data analysis, presentation/publication)  \\
    \midrule
    Novel questions & The answers to such questions are unknown and may not even exist \\
    \bottomrule
  \end{tabular}
  \caption{Two common themes defining AREd in faculty responses \cite{Redefining_authentic_research}}
  \label{tab:Definition}
\end{table*}

\begin{table*}[ht]
  \centering
  \begin{tabular}{p{0.2\textwidth}p{0.65\textwidth}}
    \toprule
    Student-to-qualified-instructor ratio & There is a need for highly qualified (in pedagogy, material knowledge, and research methods) instructors, and a low student-to-instructor ratio \\
    \midrule
    Lab/Data cost & Modern science is advanced - There is a need for advanced equipment, expensive consumables, etc. \\
    \midrule
    Skills and Knowledge & Students should have sufficient skills and knowledge to tackle AREd tasks \\
    \midrule
    Project formulation & Finding a good project may require significant time and expertise \\
    \midrule
    Time for execution & Execution of an authentic research project takes time, which often goes beyond the time ranges available in traditional courses \\
    \bottomrule
  \end{tabular}
  \caption{Barriers for wide AREd adoption at universities}
  \label{tab:blockers}
\end{table*}

\section{Ten guiding principles for AREd program design}

Here, we formulate the principles that we believe allow educators to design programs that are not limited by the barriers listed above. 

\begin{enumerate}
    \item Collaborate with researchers with expertise and data to have a basis for formulating worthy, unique projects, and who would like to explore the benefits of AI for their research
    \item Build multidisciplinary teams of students with complementary skills and perspectives: research, AI technology, management (and others when applicable). Only accept students whose existing skills can benefit a team
    \item Expect to use data provided by collaborators. Suggestions to collect new kinds of data may be provided, but don't expect students to learn and use scientific equipment to produce new data
    \item Apply best practices for work organization accepted in the industry
    \item Treat each team of students as a small startup or a consulting team working on a proof-of-concept project: timelines, priorities, success (everyone as a unit or no-one)
    \item Prioritize practical results/delivery (even if small) to long-term research ideas
    \item Organize student leadership structure to enable continuity/succession of projects and collaborations, retention of expertise, promotion of idea "we, students, have enough expertise to own and deliver", as well as streamlined management
    \item Use a "Hybrid approach to AREd": A student has exposure to and influence on all activities of the team but with a focus/priority on one of them only
    \item Keep ownership of the project within a team, with plenty of freedom for teams to direct development. Keep an eye on and don't allow your students to become "free labor" for collaborator's lab
    \item Keep a clear focus on science and adoption of new technologies in sciences (don't become an "engineering shop" for various IT-related projects)
\end{enumerate}

\subsection{A note on "Hybrid approach to AREd"}
\label{modes-of-operation}

While implementing AREd program, one can aim to provide a student with all of the most important components (table \ref{tab:Definition}) of the research process simultaneously. In such an approach, a student is expected to do everything: literature review, project formulation, experimentation, presentation of results, and more. 

Alternatively, we can expect a student to work on one small part of a large project, with only one type of activity required, such as gathering data in a lab, coding, or literature review, but not all of that together. We can see such methodology in the arrangement of common classes, where students learn math, writing, and science independently, with the assumption that a student could use all of that together later if needed.

The work of an experienced researcher is a combination of both: a researcher would work on all of the aspects of research but with a certain priority given to each of them (that priority changes over time depending on a need and an interest). As we will describe further, we employ a similar model - a "hybrid approach" in which a student has exposure to and influence on all activities of the team but with a focus/priority on one of them only (focus/priority may change from semester to semester).

As a side benefit, although an important one, the hybrid approach gives students a chance to participate in activities that can be found in real research but are usually hard to include in an AREd program if a student is supposed to focus on many aspects at once (one would need to prioritize heavily). Activities such as outreach to the "out-of-your-domain" community, writing/journalism on scientific topics, marketing and commercialization of scientific solutions, and developing scientific code into useful ready-to-consume products are, in some sense, non-essential but do allow to bring research experience out of the purely academic silo. This is especially important given that the majority of people doing research would work outside of academia after they graduate \cite{Hayter2018FactorsTI}. As educators, we need to demonstrate to students through practice that there are many ways to contribute to scientific research meaningfully, and each of them is important.

\subsection{Guiding principles in practice}

Now, let us look at the details of the implementation of the AIfSR program at NYU, which was formed with these principles in mind. This will add practical details and will allow us to see why these principles accomplish the stated goal.

\section{AIfSR group organization}

The AI for Scientific Research (AIfSR) group was formed and is operating at NYU under the umbrella of Vertically Integrated Projects\footnote{https://www.vip-consortium.org/} - VIP for short. Note that VIP can be used for various pedagogy approaches, and one can find reports on other programs relying on VIP machinery \cite{strachan2019using,coyle2006vertically}. 

\begin{figure*}
  \centering
  \includegraphics[width=1\textwidth]{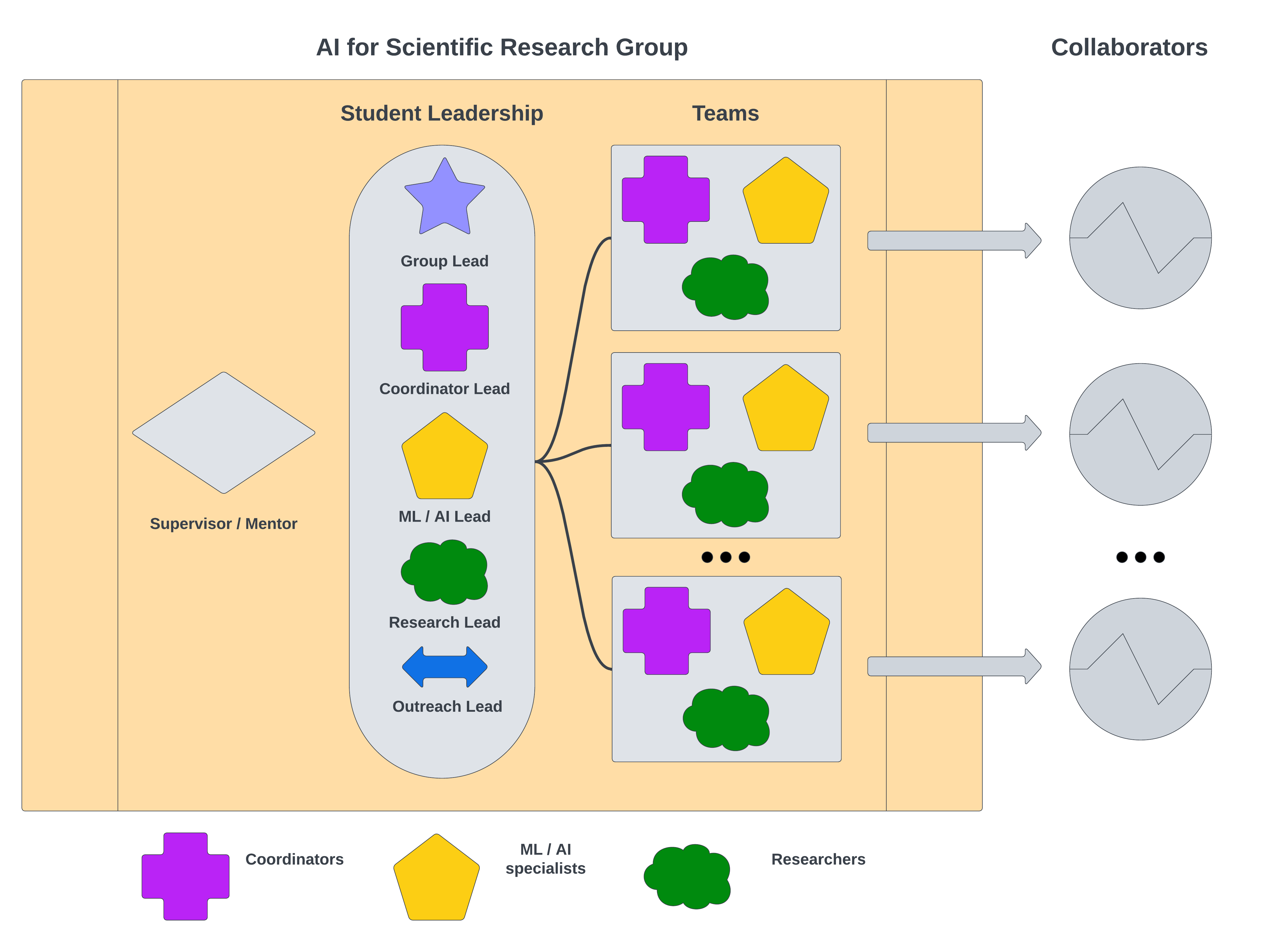}
  \caption{AI for Scientific Research (AIfSR) group structure}
  \label{fig:structure}
\end{figure*}

The AI for Scientific Research (AIfSR) group is composed of bachelor's and master's students from NYU. The whole group is divided into small student teams (usually 3-4 students), with each of them operating as a small consulting company for a collaborating lab (Fig. \ref{fig:structure}). We involve:

\begin{itemize}
    \item students who study ML (machine learning)/AI, who have previous experience with the research process in science, and those who have had experience leading others;
    \item researchers in natural sciences labs, who may have unique datasets but don't have resources (time/expertise) to develop AI;
    \item VIP program administration at university helping with outreach, enrollment process, and other related tasks; 
    \item faculty member serving as a facilitator of the group's operations;
    \item mentors who have industry experience in AI and are interested in helping/mentoring a team (or teams) of students. At present, the group has one supervisor who performs both roles: that of a mentor and of a faculty facilitator. 
\end{itemize}

\subsection{Teams and roles}

Each student in the group (outside of student leaders) is assigned to a team (Fig. \ref{fig:structure}) for the semester. Each team is tasked with tackling one project. Tasks are set up in such a way that they are often self-contained, meaning the team does not need to rely on other teams to succeed, though there is sometimes potential for collaboration between teams depending on member expertise and the similarity of the methods needed in the projects. Student leadership assigns students to groups based on their interests, skills, and preferences.  Each member of a team is assigned a role in a team. Role division brings together students with complementary skill sets that are more capable of tackling a research project.

\textbf{\textit{Coordinator role:}}
A coordinator manages the team's overall work. This role handles communication, time management, project plans for the team's activities, as well as tracking and reporting the team's progress. Coordinators are also responsible for communicating and aligning with the team's collaborators. Coordinators are expected to engage leadership more than usual if additional help or resources are required, especially if the team seems to be stuck or is moving in the wrong direction. Oftentimes, these are students who have demonstrated prior leadership experience.

\textbf{\textit{Researcher role:}}
The researcher's responsibility is to understand thoroughly the needs of the team’s collaborator. The team should clearly understand the research workflow they aim to optimize or enhance. The team needs to be able to propose how AI/ML can help in a specific laboratory. The researcher is responsible for building a strong knowledge and understanding foundation for the domain area and making sure that the solution that will be built will actually help in real life. Communicating/translating between ML and Science language is an essential part of the responsibilities. A person in the researcher role is expected to act as a scientific communicators who can survey the relevant literature and communicate the important details to other team members and any external audiences. This role is usually assigned to students with demonstrated research experience. As we have observed from participating students in past semesters, a background in the sciences is desirable, though not strictly necessary.

\textbf{\textit{ML role:}}
The role of the ML specialist is to implement the solution to the task. The ML/AI specialist is responsible for all parts of the solution, including, but not limited to, data processing, model development, model evaluation, and code management. A person in the ML role is required to have sufficient knowledge of machine learning methodologies to be able to apply them properly to any given task. This role is not meant to teach students ML from the ground up. This role provides an opportunity to apply the skills and knowledge obtained in a classroom and during work on popular datasets (such as those at Kaggle). That being said, we have many participants who come with extensive experience in industry or research and are joining the team to have an opportunity to work on really unique datasets and tasks with real-life impact.

\subsection{Management}
The group is managed by a leadership team (Fig. \ref{fig:structure}) that consists of five students and one senior supervisor. Student leadership consists of one student leader for the group, three student leaders for each role (role leads), and one outreach lead. It is possible to operate at a reduced capacity without student leadership at the initial stage.

Role leadership consists of students who act as supervisors for that given role and also interview applicants (our application system involves a centralized application portal). Typically, these are students who have displayed a high level of aptitude in their role, to the point that they would be able to guide others in said role. The outreach lead works on establishing and supporting a pipeline for the potential projects for the group and promoting the group within the university. 

The group supervisor (or people in two independent roles of facilitator and mentor) is a senior staff or faculty member who guides the group. The facilitator's responsibility includes handling any administrative tasks that student leadership does not have the authority to handle. The mentor's responsibility is to provide guidance to the group and student leadership where necessary. 

\section{AIfSR projects}

\subsection{Focus on "AI + science"}

The group's goal is to work on "AI + X" projects, where X is a specific area in one of the natural sciences. These areas have plenty of opportunities for relatively simple AI solutions to bring many benefits. Those scientific labs where AI/ML/advanced statistics is already widely used are not the targets for the teams.

We mostly work with scientific labs and not typical commercial data science companies for several reasons listed below:

\begin{itemize}
    \item This allows us to naturally include students from various fields (not only computer science students).
    \item Students in college typically don't have more than 4-8 hours per week to contribute to a program like AIfSR. Scientific projects are a great fit for the limited number of hours per week, as the timescale for project completion in academia is usually noticeably longer than in industry. 
    \item Scientists are more eager to share unique data of real value. This is rarely the case with commercial companies.
    \item Developing AI applications in the field of science helps progress the state of knowledge in the field further, which is a fulfilling experience for students.
\end{itemize}

In addition, we do not work on openly available datasets. Even a good standard ML solution developed with a unique dataset for a scientific lab is a valuable achievement. This contrasts with various projects on openly accessible data sets, where many solutions have already been implemented, and a significant amount of expertise and time commitment is required to achieve any noticeable impact. 

The principles above allow students to work on projects that impact the real world. We work with scientists whose workflow would improve immediately if a good solution is implemented.  This is in contrast to a typical situation of a non-experienced student joining a research lab only to get assigned to a small, often non-significant part of an already existing project (a similar situation is with interns in commercial companies). 

\subsection{Three stages of a project}

We found there are three distinct stages in each project. Each has its own challenges and priorities. While one may expect these activities to go in consecutive order, in reality, multiple iterations for each stage and jumps between stages are expected (as in any agile development cycle).

\begin{itemize}
    \item \textit{Formulating a project}. A crucial and important step: collaborating researchers often may not know all the opportunities AI/ML can provide. As such, teams initially spend a significant amount of time understanding the workflow in the lab. After that, members may propose and design an ML/AI-powered way to address the lab's needs. 
    \item \textit{Developing a solution}. In this stage, teams develop ML solutions, which may be based on classical or deep learning. Developed code is committed to the GitHub repository and is expected to be documented sufficiently (but not overwhelmingly so). 
    \item \textit{Delivering a solution}. A solution can make a difference only if it can be used. Our teams aim to deliver developed software to their client in a form that is easy to use. Depending on the use case, this could be as straightforward as a well-documented notebook, an interactive dashboard, and/or a R/Python package. Publishing an article or a well-written blog post describing the challenges and solutions may also be a good way to finalize the project. 
\end{itemize}

\subsection{Examples of the projects}

Here is the list, which can give a general idea about some of the projects we have worked on. 
\\

  \textit{Biology}

    \begin{itemize}
        \item Vaccination response prediction
        \item Sleep patterns classification in Electroencephalography data
        \item Optical microscope images segmentation for determination of cells' shapes
        \item Image-based prediction for worms embryonic lethality
        \item Classifying disease (malaria or not) based on symptoms
        \item Genetics: protein sequences analysis with Large Language Model (LLM)
    \end{itemize}

    \textit{Chemistry}

    \begin{itemize}
        \item Determine the electric current in batteries from the magnetic field around them
        \item Scoring and sampling for molecular crystal ranking
        \item Cancer detection through Mass Spectrometry Imaging using ML
    \end{itemize}

    \textit{Physics}

    \begin{itemize}
        \item Focused Ion Beam (FIB) beam profile determination based on etching profiles
        \item Analysis of motion trajectories of particles under a microscope
        \item ML-enhanced analysis of spectral data collected for micro-diamonds
        \item Detection and classification of pests' presence based on acoustics data
        \item AI-enabled analysis of the movement of antiferromagnetic domain walls over time
    \end{itemize}

    \textit{Psychology}

    \begin{itemize}
        \item Generative AI model for morphemes and connection to Magnetoencephalography data (Neurolinguistics of Morphology)
        \item Use of LLMs to label generic gender categories text
    \end{itemize}

    \textit{Scientific Software}

    \begin{itemize}
        \item Automatic annotation generation for scientific figures
        \item Addressing challenges of automatic parsing of scientific papers PDFs
        \item Improving benchmarking for cloud coverage detection software with a performance evaluation on HPC (High Performance Computing) cluster
    \end{itemize}

\section{AIfSR group operations}

\subsection{Weekly meetings}

Weekly meetings are required for each team to work collaboratively, exchange ideas, merge individual works, and plan for future steps. Weekly meetings usually have a flexible time length depending on the team's progress and team members’ preferences, ranging from half an hour (most of the work is completed individually) to three hours (the team prefers to work collaboratively).  

\subsection{Reporting work and progress tracking}

To effectively monitor team progress, each member of the leadership team is assigned to monitor two to three teams for close progress tracking. The team assignment is based on the development stage of teams. For example, for teams at the stage of defining a clear task, the Research Lead will be assigned to monitor this team since research-based work is intensively needed, whereas teams in the process of developing a solution would ideally be overseen by the ML lead.

\subsection{Mentor's involvement}

We found that if the mentor is involved too much, this suppresses the autonomy of a team and also prevents the leadership team from guiding teams on their own. On the other hand, if the mentor only follows progress by written records teams make and leadership reports, this leads to situations when some teams are not performing as well as they could. In our case, the balance is reached with a 15-minute catch-up
meeting with each team once in two or three weeks.

\subsection{Meetings with collaborators}

The relationship between a team and a collaborating lab resembles one between a consulting company and a client. Students of the team do not become part of the collaborating lab research group and only interact with collaborators to understand the processes in their lab better, figure out how AI/ML can help them, and then deliver and evaluate the efficiency of a solution.

While the leadership team helps teams to get started with one of the project types, teams have a lot of freedom in defining specific details of the project, specifying where the project will go and how to get there, and identifying/finding new collaborations if desired. Coordinators in each team will set up meetings with collaborators whenever the team achieves certain accomplishments or needs more information. Such meetings are infrequent (normally monthly), and at least one of the members of the leadership team will join the meeting. 

\subsection{Role specific meetings}

Role-specific meetings are held by the respective leads for all participants in that role. They happen every other week. The goal of these meetings is to build a community of researchers/machine learning engineers/coordinators, as well as boost the development of students in these roles. Unlike catch-up meetings, these meetings are less structured and less formal. This allows the students to get to know each other and increases the flow of communication.

ML role meetings may have themes such as "use of Git", "sufficiently documented code", "software useful to end user", "maintainable software", "packaging software", and "interactive dashboards".

Researcher role meetings may have themes such as "efficient reading of scientific papers, "establishing an understanding of a scientific domain", "ideation", "increasing visibility and reach of the project", and "user acceptance testing".

Coordinator role meetings may have themes such as "understanding the team's strengths and weaknesses", "establishing realistic goals", "tracking the progress", "dealing with difficulties in teams", and "career development".

For all role meetings, we encourage each member to prepare a 1-minute “sales pitch” for the project they are associated with. In the short elevator pitch, students should briefly introduce their project and describe what they have done. The practice of a 1-minute “sales pitch” enhances students' presentation skills and allows them to better present their work during future job interviews. It also serves as a self-assessment instrument and improves their understanding of the project and its importance.

\subsection{Group-wide presentations}

We have two group-wide presentations within a semester: a mid-term group presentation and a final group presentation. During the presentation, each team will give a five-minute formal presentation to explain their project and show their team's progress to all members of our group. After each presentation, other members are encouraged to ask questions and engage in discussion. The group presentation not only helps students organize their thoughts and practice their presentation skills but also encourages them to exchange ideas.

\subsection{Excellence recognition awards}

Based on the feedback we collect from the group members during the final presentation, we award gold, silver, and bronze recognition awards, which are meant to acknowledge the exceptional work our teams do.  Titles are based on two aspects: Leadership and Group members vote to evaluate the content of the team's work, as well as their final project presentation. We intentionally do not specify the criteria for our recognition awards and keep them subjective by design. Our members provide feedback based on their perception of what level of recognition a certain team should be given for their work. We believe that in science, it is not about being better than others - it is about doing the work of a certain quality level.

\subsection{Interviewing new members}

We build teams with complementary skills. Some students have already developed expertise in machine learning, some in understanding research processes, and some in communication and/or leadership. To make an efficient "unit" we carefully compose teams. We interview all shortlisted applicants to make sure people have sufficient expertise to serve well in one of the roles in the team and to explain to students what to expect. Some students have expectations that don't fit our model of operations. 

Current members of the team are asked to reapply every semester and have higher priority for consideration. Some students may wish to interview for other roles to expand their skill set and experience the research process from another point of view. This enforces opportunities for a hybrid type of AREd experience. 

When assigning new members to teams, the leadership team takes into account students' preferences and backgrounds. This may include such things as current major, experience in a specific science area, and/or expertise in specific techniques (deep learning, classical machine learning, time series analysis, unsupervised learning, etc.)

\subsection{Contribution to other roles}

It is expected for each role specialist to drive that specific direction of the project, as well as contribute a certain amount of effort to other roles' responsibilities to foster teamwork and great educational experience. A word of caution: it may happen that a person with more initiative/time/skills in one role would decide to do the job of a person in another role (for example, in a case when another person is doing the job slower than expected). Such "role takeover" should not be allowed to happen. 

\subsection{Evolution of roles}

Students who achieve excellent results and are interested in leadership responsibilities may serve as group leaders with significant influence on how the group operates. Such leadership roles allow students to guide/direct/help other students. 

\subsection{Tools and Practices}

We aim for students to use industry-standard tools and practices, thus creating additional side benefits: students can apply those tools/practices in their later work, research labs receive modern solutions, and mentors from the industry may help students use their regular tools/methods. Such skills are directly transferable to the industry, and there would be no need for participants to re-learn things later. 

We use: 

\begin{itemize}
    \item Microsoft Teams, for text communications
    \item Microsoft Office/SharePoint and Box for shared documents and data, respectively
    \item Microsoft OneNote to keep track of progress in a way that everybody on the team can see progress reported every week by another member of the team
    \item Zoom for weekly online meetings
    \item Agile-Waterfall hybrid methodology of development
    \item Python and R as main programming languages. With solutions delivered to the collaborating lab as a Python/R package or as a Python/R-based dashboard
\end{itemize}

\subsection{Knowledge transfer and assessment}

Given the significant flexibility of operations and variety between teams, it is important to understand how participants learn and what kind of feedback is available. While we did touch base on that above, here we present a more structured description.
\\

\textbf{Vertical knowledge and skills transfer}

\begin{itemize}
    \item To participants from experienced group supervisor(s), collaborating faculty, and industry professionals;
    \item  To participants from experienced students in leadership to all members;
    \item Between students within teams (Ml specialist to others, Research specialist to others, Coordinator specialist to others). This is encouraged by the expectation/requirement for each team member to not only perform their own role but also contribute time to the responsibilities of other roles. 
\end{itemize}

\textbf{Horizontal knowledge and skills transfer}

\begin{itemize}
    \item Role meetings allow students in the same role to learn various approaches to the common issue all of them face
    \item Leadership members learn from each other to deal with leadership challenges
\end{itemize}

\textbf{Vertical assessment }

\begin{itemize}
    \item Evaluation and feedback that comes from group supervisor
    \item Evaluation and feedback that comes from leadership members
    \item Feedback received from collaborators
\end{itemize}

\textbf{Horizontal assessment}

\begin{itemize}
    \item Interactions within each team and dependence of each team's progress on the contribution of each member. Participants naturally evaluate work of each other (and may encourage each other to do better) 
    \item Preparation, delivering, and receiving feedback for the "elevator pitch" mentioned above
    \item Feedback and quality-of-work labels given by peers during/after the group presentations
\end{itemize}

\textbf{Self assessment}

\begin{itemize}
    \item Individuals can assess their own work and productivity with the help of the journal they are keeping in OneNote. Each team member can see how the contribution of this team member influences the progress of the team. "Elevator pitch" mentioned above also contributes to the development of the habit of doing self-assessment
    \item Teams can assess how well they perform towards goals and priorities they set for themselves 
    \item Leadership members can asses the result of the work by observing teams dynamics and how their actions influence that
\end{itemize}

\section{Summary and conclusion}

\subsection{Is AIfSR an AREd-type program?}

From the information given in the manuscript, a reader can see that experience in the AIfSR group gives students an opportunity to meaningfully participate in all "process of science" activities listed in table \ref{tab:Definition} such as student-generated questions, hypothesis formulation, experimental design, data collection, data analysis, and presentation/publication. Students participate in all these activities while working with novel questions. 

Thus, we believe the presented framework principles allowed us to design an AREd-type program.

\subsection{Addressing barriers}

Let us review now how exactly we addressed the common AREd implementation barriers listed in table \ref{tab:blockers}. 
\newline

\textit{Student to qualified-instructor ratio}

Project development and management are handled by each team independently, with students of the leadership team providing support/guidance as needed. As such, even one faculty member can support many students.

In our particular case,  the student-to-instructor ratio is close to  30/1. The currently required involvement of the mentor is several hours per week. Supervisor involvement was greater during the first semesters while the group was only starting.
\newline

\textit{Lab/Data cost}

Data generation/collection is handled by research labs (using expensive equipment or by doing expensive 'field data collection'). Thus there is no need for additional resources to be allocated to a team of students. 
\newline

\textit{Skills and Knowledge}

In order to enable students with limited experience to work on real authentic research tasks and produce results of significant value, we form teams of students with a broader knowledge than that of any individual student. Such a team may be thought of as a "collective postdoc", or it can be looked at as a small consulting team. Multiple ways to get advice and help from peers, leads, and mentors contribute to success.
\newline

\textit{Project formulation}

It takes time and expertise to formulate a new project, which would be useful if solved, doable in a reasonable time considering resources, and not been addressed by someone else yet (or addressed only a few times in different labs). We rely on collaborating labs to provide us with unique data or expertise to formulate a problem. Members of a team would then formulate a way for how they think the lab's needs can be addressed with ML/AI/Statistics solutions. Such opportunities are unique enough, have a lot of 'freedom' to formulate what and how may be addressed, don't have a guarantee to be solvable, have real scientific value, and are approachable given students' combined expertise level. 
\newline

\textit{Time for execution}

We found that teams may need to work on a project for two (sometimes more) semesters to get to a meaningful contribution stage. By utilizing VIP program resources, we have an opportunity for students to enroll for multiple semesters. 

The student leadership team works to ensure the continuity of the projects in cases teams have new members (the teams do, in fact, often change due to the nature of the agile student body).

\subsection{Conclusion}

We believe that the principles outlined in this manuscript will allow similar programs to be implemented at many universities. Such programs would benefit all the parties involved: students, researchers in natural sciences, mentors, as well as faculty, and administrators at universities.

%\printbibliography

\bibliography{sn-bibliography}% common bib file


\begin{thebibliography}{15}
\ifx \bisbn   \undefined \def \bisbn  #1{ISBN #1}\fi
\ifx \binits  \undefined \def \binits#1{#1}\fi
\ifx \bauthor  \undefined \def \bauthor#1{#1}\fi
\ifx \batitle  \undefined \def \batitle#1{#1}\fi
\ifx \bjtitle  \undefined \def \bjtitle#1{#1}\fi
\ifx \bvolume  \undefined \def \bvolume#1{\textbf{#1}}\fi
\ifx \byear  \undefined \def \byear#1{#1}\fi
\ifx \bissue  \undefined \def \bissue#1{#1}\fi
\ifx \bfpage  \undefined \def \bfpage#1{#1}\fi
\ifx \blpage  \undefined \def \blpage #1{#1}\fi
\ifx \burl  \undefined \def \burl#1{\textsf{#1}}\fi
\ifx \doiurl  \undefined \def \doiurl#1{\url{https://doi.org/#1}}\fi
\ifx \betal  \undefined \def \betal{\textit{et al.}}\fi
\ifx \binstitute  \undefined \def \binstitute#1{#1}\fi
\ifx \binstitutionaled  \undefined \def \binstitutionaled#1{#1}\fi
\ifx \bctitle  \undefined \def \bctitle#1{#1}\fi
\ifx \beditor  \undefined \def \beditor#1{#1}\fi
\ifx \bpublisher  \undefined \def \bpublisher#1{#1}\fi
\ifx \bbtitle  \undefined \def \bbtitle#1{#1}\fi
\ifx \bedition  \undefined \def \bedition#1{#1}\fi
\ifx \bseriesno  \undefined \def \bseriesno#1{#1}\fi
\ifx \blocation  \undefined \def \blocation#1{#1}\fi
\ifx \bsertitle  \undefined \def \bsertitle#1{#1}\fi
\ifx \bsnm \undefined \def \bsnm#1{#1}\fi
\ifx \bsuffix \undefined \def \bsuffix#1{#1}\fi
\ifx \bparticle \undefined \def \bparticle#1{#1}\fi
\ifx \barticle \undefined \def \barticle#1{#1}\fi
\bibcommenthead
\ifx \bconfdate \undefined \def \bconfdate #1{#1}\fi
\ifx \botherref \undefined \def \botherref #1{#1}\fi
\ifx \url \undefined \def \url#1{\textsf{#1}}\fi
\ifx \bchapter \undefined \def \bchapter#1{#1}\fi
\ifx \bbook \undefined \def \bbook#1{#1}\fi
\ifx \bcomment \undefined \def \bcomment#1{#1}\fi
\ifx \oauthor \undefined \def \oauthor#1{#1}\fi
\ifx \citeauthoryear \undefined \def \citeauthoryear#1{#1}\fi
\ifx \endbibitem  \undefined \def \endbibitem {}\fi
\ifx \bconflocation  \undefined \def \bconflocation#1{#1}\fi
\ifx \arxivurl  \undefined \def \arxivurl#1{\textsf{#1}}\fi
\csname PreBibitemsHook\endcsname

%%% 1
\bibitem[\protect\citeauthoryear{Castelvecchi}{2015}]{Castelvecchi2015ArtificialIC}
\begin{barticle}
\bauthor{\bsnm{Castelvecchi}, \binits{D.}}:
\batitle{Artificial intelligence called in to tackle lhc data deluge}.
\bjtitle{Nature}
\bvolume{528},
\bfpage{18}--\blpage{19}
(\byear{2015})
\end{barticle}
\endbibitem

%%% 2
\bibitem[\protect\citeauthoryear{Chaturvedi et~al.}{2022}]{Chaturvedi2022InferenceOptimizedAA}
\begin{botherref}
\oauthor{\bsnm{Chaturvedi}, \binits{P.}},
\oauthor{\bsnm{Khan}, \binits{A.}},
\oauthor{\bsnm{Tian}, \binits{M.}},
\oauthor{\bsnm{Huerta}, \binits{E.A.}},
\oauthor{\bsnm{Zheng}, \binits{H.}}:
Inference-optimized ai and high performance computing for gravitational wave detection at scale.
Frontiers in Artificial Intelligence
\textbf{5}
(2022)
\end{botherref}
\endbibitem

%%% 3
\bibitem[\protect\citeauthoryear{Kim et~al.}{2019}]{Kim2019AIpoweredTL}
\begin{botherref}
\oauthor{\bsnm{Kim}, \binits{D.}},
\oauthor{\bsnm{Min}, \binits{Y.}},
\oauthor{\bsnm{Oh}, \binits{J.M.}},
\oauthor{\bsnm{Cho}, \binits{Y.}}:
Ai-powered transmitted light microscopy for functional analysis of live cells.
Scientific Reports
\textbf{9}
(2019)
\end{botherref}
\endbibitem

%%% 4
\bibitem[\protect\citeauthoryear{Romeu et~al.}{2021}]{Romeu2021CombinedMO}
\begin{barticle}
\bauthor{\bsnm{Romeu}, \binits{E.J.A.}},
\bauthor{\bsnm{Rivera-Fern{\'a}ndez}, \binits{J.D.}},
\bauthor{\bsnm{Roa-Tort}, \binits{K.}},
\bauthor{\bsnm{Valor-Reed}, \binits{A.}},
\bauthor{\bsnm{Escobedo}, \binits{G.}},
\bauthor{\bsnm{Fabila-Bustos}, \binits{D.A.}},
\bauthor{\bsnm{Isakina}, \binits{S.S.}},
\bauthor{\bsnm{Rosa-V{\'a}zquez}, \binits{J.M.}},
\bauthor{\bsnm{Guzman-Arriaga}, \binits{C.}}:
\batitle{Combined methods of optical spectroscopy and artificial intelligence in the assessment of experimentally induced non-alcoholic fatty liver}.
\bjtitle{Computer methods and programs in biomedicine}
\bvolume{198},
\bfpage{105777}
(\byear{2021})
\end{barticle}
\endbibitem

%%% 5
\bibitem[\protect\citeauthoryear{Ho}{2022}]{Ho2022BeyondIE}
\begin{barticle}
\bauthor{\bsnm{Ho}, \binits{K.Y.}}:
\batitle{Beyond images: Emerging role of raman spectroscopy-based artificial intelligence in diagnosis of gastric neoplasia.}
\bjtitle{Chinese journal of cancer research = Chung-kuo yen cheng yen chiu}
\bvolume{34 5},
\bfpage{539}--\blpage{542}
(\byear{2022})
\end{barticle}
\endbibitem

%%% 6
\bibitem[\protect\citeauthoryear{Asghar et~al.}{2021}]{Asghar2021AIIE}
\begin{barticle}
\bauthor{\bsnm{Asghar}, \binits{M.A.}},
\bauthor{\bsnm{Khan}, \binits{M.J.}},
\bauthor{\bsnm{Rizwan}, \binits{M.}},
\bauthor{\bsnm{Shorfuzzaman}, \binits{M.}},
\bauthor{\bsnm{Mehmood}, \binits{R.M.}}:
\batitle{Ai inspired eeg-based spatial feature selection method using multivariate empirical mode decomposition for emotion classification}.
\bjtitle{Multimedia Systems}
\bvolume{28},
\bfpage{1275}--\blpage{1288}
(\byear{2021})
\end{barticle}
\endbibitem

%%% 7
\bibitem[\protect\citeauthoryear{Yang et~al.}{2021}]{Yang2021ACS}
\begin{botherref}
\oauthor{\bsnm{Yang}, \binits{Y.}},
\oauthor{\bsnm{Truong}, \binits{N.D.}},
\oauthor{\bsnm{Maher}, \binits{C.}},
\oauthor{\bsnm{Nikpour}, \binits{A.}},
\oauthor{\bsnm{Kavehei}, \binits{O.}}:
A comparative study of ai systems for epileptic seizure recognition based on eeg or ecg.
2021 43rd Annual International Conference of the IEEE Engineering in Medicine \& Biology Society (EMBC),
2191--2196
(2021)
\end{botherref}
\endbibitem

%%% 8
\bibitem[\protect\citeauthoryear{Stevens et~al.}{2020}]{stevens2020ai}
\begin{botherref}
\oauthor{\bsnm{Stevens}, \binits{R.}},
\oauthor{\bsnm{Taylor}, \binits{V.}},
\oauthor{\bsnm{Nichols}, \binits{J.}},
\oauthor{\bsnm{Maccabe}, \binits{A.B.}},
\oauthor{\bsnm{Yelick}, \binits{K.}},
\oauthor{\bsnm{Brown}, \binits{D.}}:
Ai for science: Report on the department of energy (doe) town halls on artificial intelligence (ai) for science.
Technical report,
Argonne National Lab.(ANL), Argonne, IL (United States)
(2020)
\end{botherref}
\endbibitem

%%% 9
\bibitem[\protect\citeauthoryear{Gormally et~al.}{2009}]{Gormally2009EffectsOI}
\begin{barticle}
\bauthor{\bsnm{Gormally}, \binits{C.L.}},
\bauthor{\bsnm{Brickman}, \binits{P.}},
\bauthor{\bsnm{Hallar}, \binits{B.}},
\bauthor{\bsnm{Armstrong}, \binits{N.}}:
\batitle{Effects of inquiry-based learning on students' science literacy skills and confidence.}
\bjtitle{The International Journal for the Scholarship of Teaching and Learning}
\bvolume{3},
\bfpage{16}
(\byear{2009})
\end{barticle}
\endbibitem

%%% 10
\bibitem[\protect\citeauthoryear{Rodenbusch et~al.}{2016}]{Rodenbusch2016EarlyEI}
\begin{botherref}
\oauthor{\bsnm{Rodenbusch}, \binits{S.E.}},
\oauthor{\bsnm{Hernandez}, \binits{P.R.}},
\oauthor{\bsnm{Simmons}, \binits{S.L.}},
\oauthor{\bsnm{Dolan}, \binits{E.L.}}:
Early engagement in course-based research increases graduation rates and completion of science, engineering, and mathematics degrees.
CBE Life Sciences Education
\textbf{15}
(2016)
\end{botherref}
\endbibitem

%%% 11
\bibitem[\protect\citeauthoryear{Walkington and Ommering}{2022}]{Walkington2022HowDoesEngaging}
\begin{barticle}
\bauthor{\bsnm{Walkington}, \binits{H.}},
\bauthor{\bsnm{Ommering}, \binits{B.}}:
\batitle{How does engaging in authentic research at undergraduate level contribute to student well-being?}
\bjtitle{Studies in Higher Education}
\bvolume{47}(\bissue{12}),
\bfpage{2497}--\blpage{2507}
(\byear{2022})
\doiurl{10.1080/03075079.2022.2082400}
\end{barticle}
\endbibitem

%%% 12
\bibitem[\protect\citeauthoryear{Spell et~al.}{2014}]{Redefining_authentic_research}
\begin{barticle}
\bauthor{\bsnm{Spell}, \binits{R.M.}},
\bauthor{\bsnm{Guinan}, \binits{J.A.}},
\bauthor{\bsnm{Miller}, \binits{K.R.}},
\bauthor{\bsnm{Beck}, \binits{C.W.}}:
\batitle{Redefining authentic research experiences in introductory biology laboratories and barriers to their implementation}.
\bjtitle{CBE life sciences education}
\bvolume{13}(\bissue{1}),
\bfpage{102}--\blpage{110}
(\byear{2014})
\doiurl{10.1187/cbe.13-08-0169}
\end{barticle}
\endbibitem

%%% 13
\bibitem[\protect\citeauthoryear{Hayter and Parker}{2018}]{Hayter2018FactorsTI}
\begin{botherref}
\oauthor{\bsnm{Hayter}, \binits{C.S.}},
\oauthor{\bsnm{Parker}, \binits{M.A.}}:
Factors that influence the transition of university postdocs to non-academic scientific careers: An exploratory study.
Environment for Innovation eJournal
(2018)
\end{botherref}
\endbibitem

%%% 14
\bibitem[\protect\citeauthoryear{Strachan et~al.}{2019}]{strachan2019using}
\begin{botherref}
\oauthor{\bsnm{Strachan}, \binits{S.M.}},
\oauthor{\bsnm{Marshall}, \binits{S.}},
\oauthor{\bsnm{Murray}, \binits{P.}},
\oauthor{\bsnm{Coyle}, \binits{E.J.}},
\oauthor{\bsnm{Sonnenberg-Klein}, \binits{J.}}:
Using vertically integrated projects to embed research-based education for sustainable development in undergraduate curricula.
International Journal of Sustainability in Higher Education
(2019)
\end{botherref}
\endbibitem

%%% 15
\bibitem[\protect\citeauthoryear{Coyle et~al.}{2006}]{coyle2006vertically}
\begin{bchapter}
\bauthor{\bsnm{Coyle}, \binits{E.}},
\bauthor{\bsnm{Allebach}, \binits{J.}},
\bauthor{\bsnm{Krueger}, \binits{J.}}:
\bctitle{The vertically integrated projects (vip) program in ece at purdue: fully integrating undergraduate education and graduate research}.
In: \bbtitle{2006 Annual Conference \& Exposition},
pp. \bfpage{11}--\blpage{1336}
(\byear{2006})
\end{bchapter}
\endbibitem

\end{thebibliography}

\end{document}